\documentclass[journal,12 pt,onecolumn,draftcls]{IEEEtran}

\usepackage{cite}
\usepackage{amsmath}
\usepackage{amssymb}
\usepackage{graphicx}
\usepackage{graphics}
\usepackage{dsfont}
\usepackage{verbatim} 
\usepackage{algorithm}
\usepackage{algorithmic}
\usepackage{epstopdf}

\usepackage[top=0.8in, bottom= 1.1in, left=0.835in, right=0.835in]{geometry}

\begin{document}
\title{Low-Complexity Energy-Efficient Broadcasting in One-Dimensional Wireless Networks}

\author{
\IEEEauthorblockN{Mohammad R. Ataei, Amir H. Banihashemi and Thomas Kunz}
\\ \IEEEauthorblockA{Systems and Computer Eng. Dept., Carleton University, Ottawa, ON, Canada
\\
 E-mails: \{mrataei, ahashemi, tkunz\}@sce.carleton.ca}
}
\maketitle
\vspace{-1 cm}
\begin{abstract}
In this paper, we investigate the transmission range assignment for $N$ wireless nodes located on a line (a linear wireless network) for broadcasting data from one specific node to all the nodes in the network with minimum energy. Our goal is to find a solution that has low complexity and yet performs close to optimal. We propose an algorithm for finding the optimal assignment (which results in the minimum energy consumption) with complexity $\mathcal{O}(N^2)$. An approximation algorithm with complexity $\mathcal{O}(N)$ is also proposed. It is shown that, for networks with uniformly distributed nodes, the linear-time approximate solution obtained by this algorithm on average performs practically identical to the optimal assignment. Both the optimal and the sub-optimal algorithms require the full knowledge of the network topology and are thus centralized. We also propose a distributed algorithm of negligible complexity, i.e., with complexity $\mathcal{O}(1)$, which only requires the knowledge of the adjacent neighbors at each wireless node.  Our simulations demonstrate that the distributed solution on average performs almost as good as the optimal one for networks with uniformly distributed nodes.

\end{abstract}


\IEEEpeerreviewmaketitle

\section{Introduction}
\IEEEPARstart{T}{he} \emph{Minimum-Energy Broadcasting} problem in wireless networks focuses on finding a transmission range assignment for all the nodes in the network such that the total consumed energy for broadcasting data from one specific node to all the other nodes is minimized. This problem is known to be NP-hard for D-dimensional spaces with $D \geq 2$, \cite{np_hard}, \cite{Clementi_2_3_D}. In this paper, we focus on the case where the nodes are located on a line. One application of linear networks is the wireless communication in a vehicular ad hoc network (VANET), where the data transmission is along a road \cite{vanet}. In this work, we are interested in finding low-complexity solutions to this broadcasting problem which are optimal or close to optimal in terms of energy consumption. Since nodes can adjust their transmission power and therefore their transmission range, the problem is to find a transmission range assignment for all the nodes such that the source node can transmit its data to all the other nodes in the network and the total consumed energy is minimized.

Previous works \cite{Piret}, \cite{santi}, studied the problem of linear network connectivity under the assumption that the nodes are uniformly and independently distributed, and that all the nodes have the same transmission range. In \cite{Piret}, it is shown that for a linear network of length $L$, where the density of the uniformly distributed nodes is $\lambda$, the network will be connected with probability one as the length of the network goes to infinity, if all the nodes have an identical transmission range $r$ greater than $\frac{\ln(\lambda L)}{\lambda}$. Santi \emph{et al.} \cite{santi} provided both upper and lower bounds for the identical transmission range $r$ under the same network assumptions. They demonstrated that if $rN$ is $\Omega(L\ln(L))$, where $N$ is the number of the nodes in the network of length $L$, then with probability one the network will be connected.

Recently, in \cite{arbitrary_non_vanishing}, an arbitrary node distribution was considered, and it was shown that for a continuous density function $f(x)$, where the number of nodes in the network, $N$, goes to infinity,
$\tau_{f,N}=\frac{1}{\inf\{f(x)\}}\frac{\log(N)}{N}$ is the critical transmission range, where $inf\{\cdot\}$ denotes the infimum value.
This means that if every node in the network has a transmission range greater than $\tau_{f,N}$, the network will be connected with probability one. It was further shown in \cite{arbitrary_vanishing} that for the networks with vanishing density functions a strong threshold does not exist.

Non-asymptotic results for a linear network of $N$ nodes with density $\lambda$ and with exponentially distributed distances between the nodes are also available~\cite{Prob_Connectivity}, which indicate that the network will be connected with a probability greater than $P_c$, if $r \geq \frac{-\ln(1-P_c^{1/(N-1)})}{\lambda}$.
Given the node density $\lambda$ and the network length $L$, the value of $r$ can thus be determined for a value of $P_c$ arbitrarily close to one.

In \cite{artimy}, the nodes, which are considered to be vehicles, perform a distributed algorithm to estimate the local density of the nodes in the network. The algorithm uses the mobility pattern of a node (vehicle) and is based on the stopping time of that node. The nodes adjust their transmission ranges according to the estimated density in a manner similar to that of \cite{santi}.

The most relevant study to our paper is \cite{new_results}, in which the authors solve the problem of finding a range assignment for $N$ nodes in a linear network for broadcasting from a specific node to all the other nodes with minimum energy. In \cite{new_results}, the nodes' locations on the line are arbitrary and the source node is assumed to be known. It is also assumed that the network topology is known and available for solving the problem. The algorithm proposed in \cite{new_results} has complexity $\mathcal{O}(N^3).$\footnote{By definition, $f(x) = \mathcal{O}(g(x))$ if and only if there exists a positive constant $a$ such that for all sufficiently large values of $x$, we have $|f(x)|\leq a|g(x)|$.}

More recently, there have been some studies \cite{Clementi_1_D}, \cite{Clementi_weighted} focusing on the same problem with additional constraints. In \cite{Clementi_1_D}, the condition that all the nodes have to receive data in at most $h$ hops has been added, and the proposed algorithm has complexity $\mathcal{O}(hN^2)$. In \cite{Clementi_weighted}, the same problem is solved with the assumption that the consumed energy in each node depends not only on the transmission range of the node but also on an arbitrary positive weight assigned to the node. The problem is solved with algorithms of complexity $\mathcal{O}(N^3)$ for the unconstrained case (i.e., $h=N-1$) and $\mathcal{O}(hN^4)$ for the $h$-hop constrained case.

Papers \cite{Clementi_2_3_D}, \cite{Kranakis_strong}, and \cite{Das_linear_strong} studied the range assignment when the network graph is strongly connected.\footnote{A network graph is strongly connected if each node of the graph is connected to every other node in the graph.} This assignment is called a \emph{complete range assignment}. The study in \cite{Clementi_2_3_D} focused on $D$-dimensional networks with $D \geq 2$, where every path between any two nodes consists of at most $h$ hops. In \cite{Kranakis_strong}, Kirousis \emph{et al.} presented a dynamic programming algorithm with $\mathcal{O}(N^4)$ time complexity for finding a minimum cost complete range assignment, where the nodes are located on a line and the distances between the nodes are arbitrary. In \cite{Das_linear_strong}, a minimum cost complete range assignment solution was found for linear networks with an algorithm with complexity $\mathcal{O}(N^3)$, improving the time complexity of the algorithm of \cite{Kranakis_strong} by a factor of $N$.

In this paper, we tackle the energy-efficient broadcasting problem under different assumptions for the available knowledge about the network topology. In our formulation, we pose no limit on the number of hops and assume that the consumed energy by each node is a function of the node's transmission range. The system model and the problem formulation is described in Section II. In Section III, we propose our algorithms for energy-efficient broadcasting. The first algorithm finds a sub-optimal solution to the problem with complexity $\mathcal{O}(N)$. The second algorithm finds an optimal solution and has complexity $\mathcal{O}(N^2)$. Note that both algorithms are significantly less complex than the algorithm of \cite{new_results} which has a complexity $\mathcal{O}(N^3)$. Similar to the algorithms of \cite{Clementi_2_3_D}, \cite{new_results} - \cite{Das_linear_strong}, both proposed algorithms require the full knowledge of the network topology and are thus centralized. Finally, the last algorithm proposed in this work is a distributed one with negligible complexity of $\mathcal{O}(1)$. To be implemented in each node of the network, this algorithm requires only the knowledge of the adjacent neighbor(s) of the node. This is much easier to attain compared to the amount of knowledge required for the centralized algorithms, i.e., the full knowledge of the network topology. Furthermore, this simplifies the implementation of the algorithm in mobile scenarios, where nodes only need to track their two closest neighbors. 
In Section IV, we present simulation results on the performance of the proposed algorithms. Finally, conclusions are drawn in Section V.

\section{System model}
We consider a set of $N$ nodes placed on a line having indices $\{1,2,\ldots,N\}$ from left to right, and a specific node with index $s$ among them as the source. The source node $s$ broadcasts data to all the other nodes in the network. This is to be performed in an energy-efficient multi-hop fashion. To solve this broadcasting problem, we need to assign a \emph{transmission range} to each node so that the total consumed energy is minimized. A node is assumed to have symmetric coverage on both sides up to its transmission range and any other node located in the transmission range of this node can receive the transmitted data.
In this work, similar to \cite{Clementi_2_3_D}, \cite{Piret} - \cite{Das_linear_strong}, we do not consider the effects of interference caused by wireless communication among the nodes, and shadowing and fading, and also the overhead of obtaining information about the nodes' locations.\footnote{Ignoring the interference can be justified by assuming that a scheduling scheme would ensure that simultaneous interfering transmissions will not occur. For ignoring fading/shadowing, one can assume that the transmissions occur in an environment with no obstacles, where the signals experience negligible fading/shadowing. If fading/shadowing is not ignored, then each link between different nodes of the network would experience a different loss and thus a different relationship between the distance and the consumed energy. This will change the system model compared to the one discussed in this work and is beyond the scope of this correspondence.}

A \emph{range assignment} $R$ is a function $R:\{1,\ldots,N\} \rightarrow \mathds{R}^+$, where $R(i)$ is the assigned transmission range to node $i$. We denote the consumed energy of the range assignment $R$ by $cost(R)$ and assume that it can be calculated, up to a constant multiplicative factor, using the following equation:
\vspace{-0.2 cm}
\begin{align}
cost(R)=\sum_{k=1}^N{R^\alpha(k)},\nonumber
\end{align}
where $\alpha$ is the \emph{path-loss exponent} whose value is normally between 2 and 6 \cite{alpha}.
By using the \emph{Minimum Energy Range Assignment} (denoted by $R_{opt}$), every node in the network will receive the data transmitted by the source node with the minimum possible cost.

\section{Energy-Efficient Range Assignments}
In a linear network, each node has at most two immediate neighboring nodes, one on each side. We call these two nodes the adjacent neighbors of a node. In the sequel, we refer to the adjacent neighbor that is further away from the source as \emph{the next adjacent neighbor} of a node.

Let us first consider a case where the source node is on one end of the network. There is a trivial optimal solution for this case \cite{new_results}. In the optimal solution, the transmission range of each node, except the node which is at the other end of the network with respect to the source node, is equal to the distance to its next adjacent neighbor. For example if $s=1$, the optimal solution will be as following:
\vspace{-0.3 cm}
\begin{align}
R_{opt}(i)=d(i,i+1) \hskip 0.8em \text{for}\hskip 0.4em i=1,\ldots,N-1  \hskip 0.4em \text{and} \hskip 0.4em  R_{opt}(N)=0,\nonumber
\end{align}
where $d(i,i+1)$ is the distance between nodes $i$ and $i+1$.

This result is obtained by using the fact that for any given set $\{a_1,a_2,\cdots,a_W\}$ of positive numbers, where $W$ is an arbitrary integer, and for any  $\alpha \geq 2$, we have:
\begin{align}
 \left ( \sum_{k=1}^{W}{a_k}\right )^\alpha \geq \sum_{k=1}^{W}{a_k^\alpha}.
\end{align}

Hence, in the rest of the paper, we assume that the source node is not at one end of the network ($s\neq 1,N$). Since nodes $1$ and $N$ do not have next adjacent neighbors to send data to, their transmission range in the optimal solution will always be zero.

We divide the whole set of nodes excluding nodes $1$ and $N$ into two sets $\mathbb{L}$ and $\mathbb{R}$, where:
\vspace{-0.2 cm}
\begin{align}
\mathbb{L}=\{i:1< i\leq s\}  \hskip 0.4em \text{and} \hskip 0.4em  \mathbb{R}=\{i':s\leq i'< N\}. \nonumber
\end{align}
We denote $s$ by $s_L$ when it is in $\mathbb{L}$, and by $s_R$ when it is in $\mathbb{R}$.

The following lemma forms the basis of the proposed algorithms. Its proof is by contradiction and straight-forward.

\textbf{Lemma 1:}
In the minimum-energy range assignment, the transmission range of a node $i$ is either zero or greater than or equal to the distance between $i$ and its next adjacent neighbor.

The minimum possible positive range of node $i$, $M(i)$, can be calculated as:
\vspace{-0.2 cm}
\begin{align}
M(1)=M(N)=0, \hskip 0.6em
M(i)=
\left\{
  \begin{array}{ll}
    d(i,i-1)  \text{ for }i\in \mathbb{L},\nonumber
    \\d(i,i+1) \text{ for }i\in \mathbb{R}.\nonumber
  \end{array}
\right.
\end{align}
For $i = s$, there are two values of $M(i)$ corresponding to $s_L$ and $s_R$, respectively.
\vspace{-0.4 cm}
\subsection{Sub-Optimal Range Assignment with Linear Complexity}
For networks with known topology, we can save energy by preventing some nodes from transmission. These are the nodes with receivers located in the transmission range of other nodes. Fig. 1 shows an example of this situation. Node $b$ receives the data from $s$ and as $b$ needs to transmit the data at a power level that can reach its next adjacent neighbor (node $c$), the data also reaches nodes $a$ and $d$. Hence $R(a)=R(d)=R(c)=0$.

The focus of our proposed sub-optimal algorithm is on finding the nodes that can save energy by not transmitting, while the other nodes only transmit at a power level that is needed for their next adjacent neighbors to receive the data.

The pseudo-code for this algorithm is given as Algorithm 1, that has the output $R_{sub}$ as the range assignment. In Step 2 of Algorithm 1, node $s$ has two opposite side coverage values, each corresponding to one of its roles as $s_L$ or $s_R$, i.e., $Cov(s_L)=M(s_{L})$ and $Cov(s_R)=M(s_{R})$. Note that in Step 4, $l_{R}$ or $l_{L}$ can be the source node $s$. In Step 5, the two costs correspond to two different ways of sending data from $s$ to all the nodes in the network. The $cost_R$ ($cost_L$) is for the case where some nodes on the right (left) hand side of the source receive the data from $m_L$ ($m_R$) and thus do not need to receive the data from their neighbor.

The following theorem is easy to prove.

\textbf{Theorem 1:}
The sub-optimal algorithm has a time complexity of $\mathcal{O}(N)$.

In this section, we assumed that the transmission range of a node is either equal to zero or equal to the distance to its next adjacent neighbor. For finding the optimal range assignment in the next section, we will use the fact that each node can have transmission range equal to zero or its distance to any node in the network ($N$ possible values). Among all these possible assignments, we will prove that by searching a limited space the optimal solution can be found.
\vspace{-0.5 cm}
\subsection{Optimal Range Assignment}
The following theorem states some important facts about $R_{opt}$.

\textbf{Theorem 2:}
The range assignment $R_{opt}$ satisfies the following conditions:
\begin{enumerate}
  \item There exists at most one node (denoted by $b_m$) with transmission range greater than $M(b_m)$.
  \item If $b_m$ exists, it receives the data from $s$ via the nodes in between $s$ and itself (i.e., it does not receive the data via a node on the opposite side of the source).
  \item If $b_m$ exists,
  then $R_{opt}(b_m)\geq \max(d(b_m,l_R),d(b_m,l_L))$.
\end{enumerate}

\textbf{Proof:}
\begin{enumerate}
  \item First we should note that the necessary condition for any node $i$ to have $R_{opt}(i)>M(i)$ is that by transmitting at this higher power, it must have a receiver on the opposite side of the source. Otherwise, using equation (1), another range assignment with less consumed energy can be found. The proof then follows from the same arguments made in Lemmas 3 and 4 of \cite{new_results}. This is done by substituting the concept of a \emph{root-crossing} node in \cite{new_results} with a node with $R_{opt}(i)>M(i)$ in our context, and noticing that the \emph{children} of a node in \cite{new_results} are receivers of that node in our study.

  \item This is a direct result of Lemmas 3 and 4 of \cite{new_results}, which indicate that the optimal solution contains exactly one root-crossing node, i.e., if $b_m$ exists, it is the only root-crossing node.
  \item The proof for this part is by contradiction. Suppose that node $b_m$ exists 
       and is on the left side of the source (the proof for the right side is similar).

      First, we prove that if $R_{opt}(b_m)<d(b_m,l_R)$, the assignment $R_{opt}$ cannot be optimal. There can be more than one node having $l_R$ in their transmission range. We denote the right-most node among those nodes on the left side of the source sending data to $l_R$ by $mR_L$. If $b_m$ is on the left side of $mR_L$, then since $b_m$ receives the data from $s$ via the nodes in between $s$ and itself (including node $mR_L$), node $l_R$ receives data from $mR_L$. The other possibility is for $b_m$ to be on the right side of $mR_L$ (in between nodes $s$ and $mR_L$). In this case, since $M(mR_L)\geq d(mR_L,l_R)>d(b_m,l_R)>R_{opt}(b_m)$, node $b_m$ cannot transmit to the left receiver of node $mR_L$, so node $mR_L$ still needs to transmit ($R_{opt}(mR_L)=M(mR_L)$), which implies that $l_R$ will be covered by the transmission from $mR_L$. Hence with $R_{opt}(b_m)<d(b_m,l_R)$, node $b_m$ cannot transmit data beyond node $l_R$ on the right side of the source. Since all the nodes from $s$ to $l_R$ can receive the data from $mR_L$, node $b_m$ has no receivers on its opposite side. Using equation (1), we can easily show that by using the sub-optimal range assignment $R_{sub}$, all the nodes can still receive the data with a lower cost, contradicting the optimality of the range assignment.

      Now suppose that $R_{opt}(b_m)<d(b_m,l_L)$. Note that since $d(b_m,l_R)\geq d(b_m,s)$, if node $b_m$ is on the left side of node $l_L$, we will have: $R_{opt}(b_m)<d(b_m,l_L)\leq d(b_m,s) \leq d(b_m,l_R)$ which results in a contradiction as previously discussed. We continue the proof for the case that node $b_m$ is in between nodes $s$ and $l_L$. Denote the left-most node on the right side of the source having $l_L$ in its transmission range by $mL_R$. Also denote the last same-side receiver of node $mL_R$ by $ls_m$ (which is on the right side of the source). Since $d(b_m,ls_m)>M(mL_R)=d(mL_R,ls_m)\geq d(mL_R,l_L) > d(b_m,l_L) >R_{opt}(b_m)$, node $b_m$ cannot transmit to node $ls_m$, so node $mL_R$ still needs to transmit ($R_{opt}(mL_R)=M(mL_R)$). Node $mL_R$ is a root-crossing node, and since there cannot be two root-crossing nodes in the optimal solution, node $mL_R$ does not receive the data via a node on the other side of the source (e.g., node $b_m$). Therefore node $b_m$ has no receivers on the opposite side, and it has no receivers beyond node $l_L$, and since all the nodes in between nodes $s$ and $l_L$ receive the data from node $mL_R$, we can assign zero to the transmission range of node $b_m$. This results in reducing the energy consumption of $R_{opt}$, which contradicts its optimality.

      We showed that $R_{opt}(b_m)\nless d(b_m,l_R)$ and $R_{opt}(b_m)\nless d(b_m,l_L)$, therefore \\ $R_{opt}(b_m)\geq \max(d(b_m,l_R),d(b_m,l_L))$, and this completes the proof.
      \vspace{-0.75 cm}
      \flushright{$\Box$}

\end{enumerate}
The following corollary is obtained based on the second and third propositions of Theorem 2.

\textbf{Corollary 1:}
If node $b_m$ exists in $R_{opt}$, then
\begin{align}
cost(R_{opt})= \nonumber
\left\{
  \begin{array}{ll}
    R_{opt}^\alpha(b_m)+E_L+E_R & \hbox{if $b_m=s$,} \\
    \sum_{k=b_m}^{s-1}{d^\alpha(k,k+1)}+R_{opt}^\alpha(b_m)+E_L+E_R & \hbox{if $1<b_m<s$,} \\
    \sum_{k=s}^{b_m-1}{d^\alpha(k,k+1)}+R_{opt}^\alpha(b_m)+E_L+E_R  & \hbox{if $s<b_m<N$,}
  \end{array}
\right.
\end{align}
where $E_L=\sum_{j=1}^{r_L-1}{d^\alpha(j,j+1)}$ and $E_R=\sum_{j'=r_R}^{N-1}{d^\alpha(j',j'+1)}$ in which  $r_L$ and $r_R$ denote the last left-side and right-side receivers of node $b_m$, respectively.

For reducing the complexity of the optimal algorithm, we introduce two arrays and one matrix as follows. Let $C_S$ be an array of size $N+1$, where $C_S[i]$ ($i=1,\ldots,s_L,s_R,\ldots,N$) is the cost of sending data from node $s$ to node $i$ via the nodes in between them. Construction of this array is performed by the following recursive equations, having time complexity of $\mathcal{O}(N)$.
\begin{align}
&C_S[s_L]=C_S[s_R]=0,\hskip 0.4em \nonumber
&C_S[i]=
\left\{
  \begin{array}{ll}
    C_S[i+1]+M^\alpha(i+1) & 1\leq i < s_L, \\ \nonumber
    C_S[i-1]+M^\alpha(i-1) & s_R< i \leq N. \nonumber
  \end{array}
\right.
\end{align}

Similarly, we can compute another array $C_E$ of size $N+1$, where $C_E[i]$ ($i=1,\ldots,s_L,s_R,\ldots,N$) is the cost of sending data from node $i$ to the end node on its side, via the nodes in between them. Using the following recursive equations, we can construct $C_E$ with complexity $\mathcal{O}(N)$ in time.
\vspace{-0.2 cm}
\begin{align}
&C_E[1]=C_E[N]=0, \hskip 0.4em \nonumber
&C_E[i]=
\left\{
  \begin{array}{ll}
    C_E[i-1]+M^\alpha(i) & 1< i \leq s_L, \\ \nonumber
    C_E[i+1]+M^\alpha(i) & s_R\leq i < N. \nonumber
  \end{array}
\right.
\end{align}

Denote by $LR$ a $2\times N$ matrix in which column $i$ contains the indices of the last same-side and other-side receivers of node $i$ (denoted by $rS_i$ and $rO_i$, respectively), when the transmission range of node $i$ is equal to $\max(d(i,l_L),d(i,l_R))$. Note that constructing this matrix has a time complexity of $\mathcal{O}(N^2)$.

The pseudo-code for the optimal algorithm is given in Algorithm 2. The algorithm has the output $R_{opt}$ as the optimal range assignment.

\textbf{Theorem 3:}
Obtaining $R_{opt}$ by Algorithm 2 has complexity $\mathcal{O}(N^2)$.

\vspace{-0.4 cm}
\subsection{Distributed Range Assignment}
In the last algorithm, every node just knows the distances to its adjacent neighbors. The amount of required information in this case is much less than that of the previous scenarios. Since each node only requires local information, this algorithm can be implemented in a distributed manner.

In the proposed distributed algorithm, each node should wait till it receives data from one of its adjacent neighbors. It then transmits the data to its other neighbor. The range assignment $R_{adj}$ for this algorithm is thus:
\begin{align}
 R_{adj}(s)=\max\{M(s_R),M(s_L)\} \hskip 0.4em \text{and} \hskip 0.4em R_{adj}(i)=M(i) \text{ for } i\neq s. \nonumber
\end{align}

If we assume that the $N$ nodes are located on a line according to a uniform distribution, and independent of one another, then the distances between the adjacent nodes are independent and identically distributed (i.i.d.) random variables $\{D_1,...,D_{N-1}\}$, where $D_i$ denotes the distance between nodes $i$ and $i+1$. These random variables have an exponential distribution, denoted by $exp(\lambda)$ with the probability density function given by $\lambda e^{-\lambda x}$ for $x \geq 0$, where $\lambda$ is the density of the nodes on the line. Under such assumptions, the expected total cost of the distributed range assignment can be calculated as follows.
\vspace{-0.2 cm}
\begin{align}
\text{\emph{Expected Cost}}=\sum_{\substack{k=1\\k\neq s-1,s}}^{N-1}{E[(D_k)^\alpha]}+E[(\max\{D_{s-1},D_s\})^\alpha]=\frac{\alpha!}{\lambda^\alpha}\left ( N-1- \frac{1}{2^\alpha}\right ).
\end{align}

\subsection{Identical Transmission Range}
When the distances between the nodes are drawn i.i.d. from $exp(\lambda)$, a simple solution to the transmission range assignment, for maintaining the network connectivity with a given probability $P_c$, is to assign an identical transmission range $R(P_c)$ to all the nodes in the network.  This assignment must satisfy the following inequality:
\vspace{-0.2 cm}
\begin{align}
R(P_c)\geq \frac{-\ln(1-P_c^{1/(\lambda L -1)}
)}{\lambda}.\nonumber
\end{align}

Note that this equation is similar to the one given in \cite{Prob_Connectivity} where $L$ is approximated by $N/\lambda$. By ignoring the term -1 in $\lambda L - 1$, and using the first two terms of the Taylor series of $P_c^{1/(L \lambda)}$ for variable $1/L$ in the neighborhood of zero, we obtain the following approximation for the lower bound of the identical transmission range:
\vspace{-0.2 cm}
\begin{align}
R_l(P_c) \approx \frac{\ln(\frac{-\lambda L}{\ln(P_c)})}{\lambda}.
\end{align}

The range assignment in (3) can be used when the only available information about the network is its density and length. Moreover, using this range assignment, there is no guarantee that the network will in fact be connected. On the other hand, the condition of the network being connected is in general stronger than the condition required for a specific source node in the network to broadcast its message to all the other nodes in the network.
\vspace{-0.3 cm}
\section{Numerical Results}
The simulation results are presented for networks with $N$ nodes distributed uniformly and independently over a line of length $L$. In simulations, we assume $\lambda = N/ L$ and $\alpha=2$. For each simulation point corresponding to a given density $\lambda$, $10000$ random networks are generated. For the results comparing the different algorithms, we run the algorithms on exactly the same networks and obtain the average of the total consumed energy over the $10000$ networks.

In Fig. 2, we compare the energy consumption of the identical range assignment of Section III.D (for different $P_c$ values) with those of the sub-optimal range assignment of Section III.A, the optimal range assignment of Section III.B and the distributed range assignment given in Section III.C for a network with $L = 5000$ meters. As can be seen in Fig. 2, in general, the total consumed energy decreases as the density of the nodes increases. This can be explained by inequality (1), where nodes with closer distance to each other can communicate with less energy over a given distance compared to nodes that are further apart. The proposed assignments significantly outperform the identical range assignment, even for a $P_c$ value as small as 0.85. Interestingly, both the linear-time sub-optimal algorithm and the distributed algorithm perform practically the same as the optimal algorithm over the whole range of network densities. For the distributed algorithm, the simulation results and the analytical results from (2) are almost identical.

To obtain a more detailed picture of the relative energy consumption of the proposed algorithms, we have plotted the histogram of the normalized difference between their consumed energy for $\lambda = 0.03$ in Fig. 3. The normalized difference between the energy consumption of range assignments $R_1$ and $R_2$  is defined as:
\vspace{-0.15 cm}
\begin{equation}
\label{normalized}
\frac{\max(cost(R_1),cost(R_2)) - \min(cost(R_1),cost(R_2))}{\min(cost(R_1),cost(R_2))}\:.
\end{equation}

Fig. 3 demonstrates that the normalized difference between the energy consumption of the three proposed algorithms is rather small (less than $10\%$) for the simulated cases. The simple distributed algorithm and the linear sub-optimal algorithm consume at most $9\%$ and $6\%$ more energy than the complex optimal algorithm, respectively.

Although the difference between the energy consumption of the more complex range assignment algorithms compared to less complex ones is rather small in networks with uniformly distributed nodes, there exist network topologies where such differences are large. Fig. 4 illustrates two examples of such topologies. In both examples, we assume that $\epsilon_1, \epsilon_2 \ll r_1, r_2$, and that $\epsilon_1 \simeq \epsilon_2$, for simplicity. In the network given in Fig. 4(a), assuming that $r_1\geq r_2+\epsilon_1+\epsilon_2$, and $r_1 \simeq r_2$, the normalized difference in energy consumptions of the linear sub-optimal and distributed algorithms, calculated by (\ref{normalized}), is about $100\%$.

In the network given in Fig. 4(b), assuming that $r_1\leq r_2+\epsilon_1+\epsilon_2$ and $r_1+\epsilon_1\geq r_2+\epsilon_2$, in the optimal range assignment, node $s$ sends data to node $c$, that in turn transmits data to nodes $a$, $b$ and $d$. In the linear sub-optimal range assignment, node $s$ transmits data to node $d$, and thus also covers nodes $c$ and $b$. Node $b$ then transmits data to node $a$. Assuming $r_1 \simeq r_2$, the normalized difference between the consumed energy of the optimal and the linear sub-optimal algorithms is about $100\%$.

Finally in Fig. 5, we show the histogram of the distance $d(b_m,s)$ between node $b_m$ in the optimal solution and the source node for $10000$ generated networks with $\lambda=0.03$. This corresponds to the same scenarios used for Figs. 2 and 3. The histogram of Fig. 5 shows that for the majority of cases, node $b_m$ does not exist. In most of the remaining cases, the distance $d(b_m,s)$ is relatively small, with the maximum distance less than $30\%$ of the length of the network. These are the main reasons behind the small difference among the energy consumption of the three proposed range assignments in the simulated scenarios as reflected in Figs. 2 and 3. A careful inspection of Figs. 3 and 5 also demonstrates that although there exist some cases where  $d(b_m,s)$ is relatively large (about $0.3L$), the difference in energy consumption is relatively low (less than $10\%$). The reason is that in such cases although energy is saved in the optimal solution through the coverage of opposite side nodes by node $b_m$, node $b_m$ itself consumes a large amount of energy due to its large transmission range.

\vspace{-0.3 cm}
\section{Conclusion}
In this work, we proposed solutions for energy-efficient broadcasting in linear networks. Two of these solutions are centralized algorithms. One finds the exact transmission range assignment for minimum-energy broadcasting, and has complexity $\mathcal{O}(N^2)$, where $N$ is the number of network nodes. This improves the complexity $\mathcal{O}(N^3)$ of existing solutions. The other centralized algorithm is linear-time and finds an approximation of the optimal solution. Furthermore, we proposed a simple distributed range assignment algorithm for energy-efficient broadcasting. We demonstrated that on average both the linear-time approximation and the distributed algorithm are almost as efficient as the optimal range assignment for networks with uniformly distributed nodes. The distributed algorithm would be of particular interest not only because of its distributed nature, but also for its very low complexity (constant in network size), and the small amount of network knowledge that each node requires to perform the algorithm (only the distances to the adjacent neighbors).

\section*{Acknowledgment}
The authors wish to thank the anonymous reviewers whose comments improved the presentation of the paper.

\vspace{-0.3 cm}
\bibliographystyle{IEEEtran} 

\bibliography{refs}  

\newpage

\begin{figure}[h!]

  \centering
    \includegraphics[width=6in]{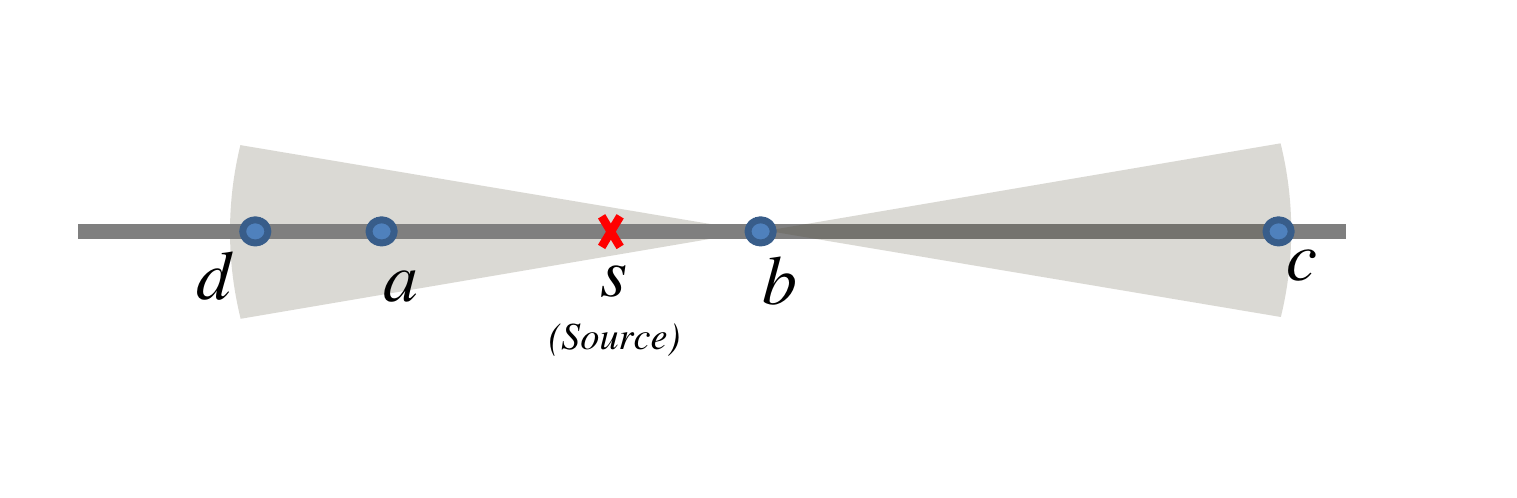}
     \caption{In an energy-efficient range assignment, some nodes do not need to transmit.}
\end{figure}

\begin{figure}[h!]

  \centering
    \includegraphics[width=5.5in]{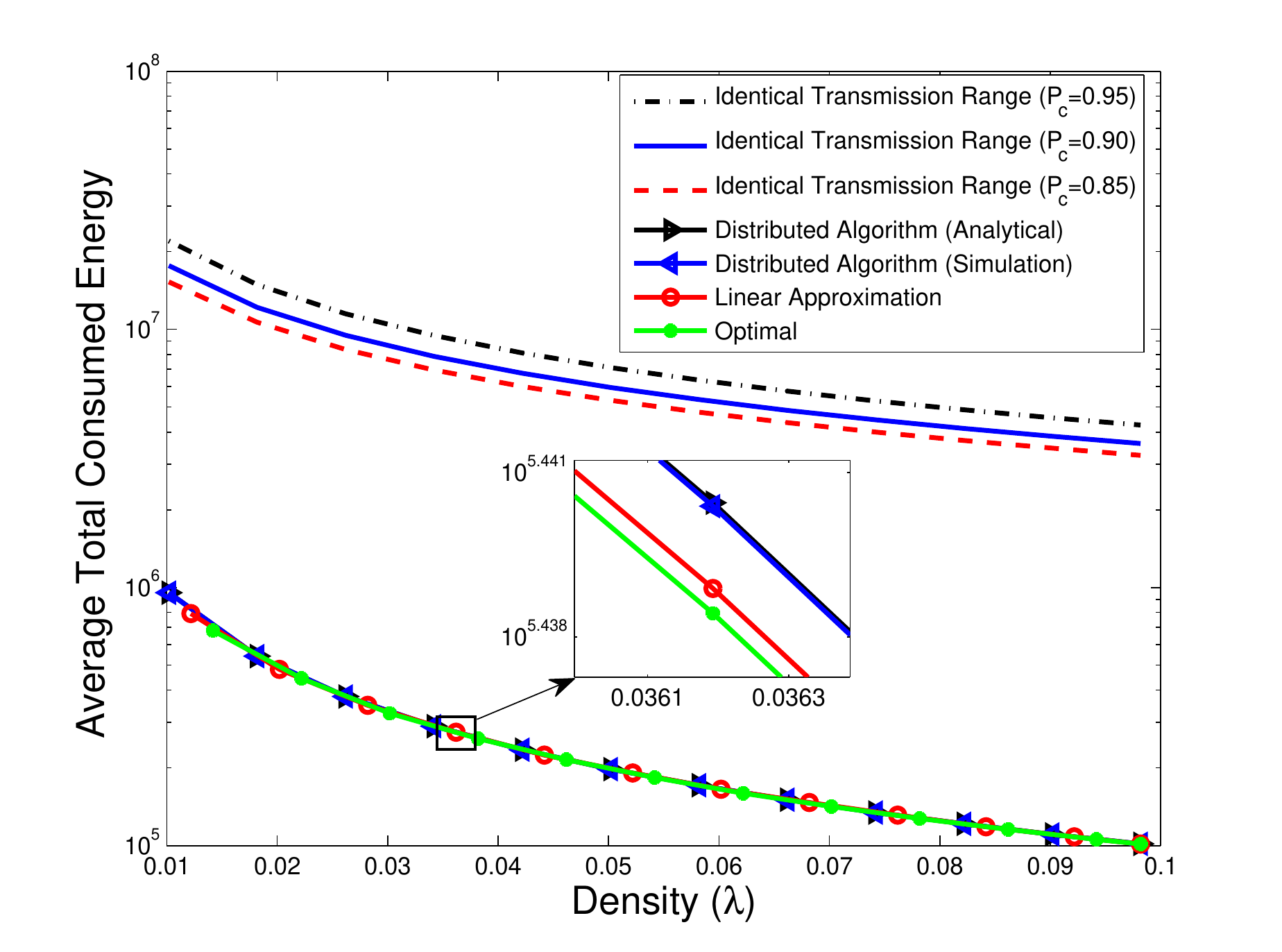}
     \caption{Comparison of the energy consumption of different range assignments.}
\end{figure}

\begin{figure}[h!]

  \centering
    \includegraphics[width=6in]{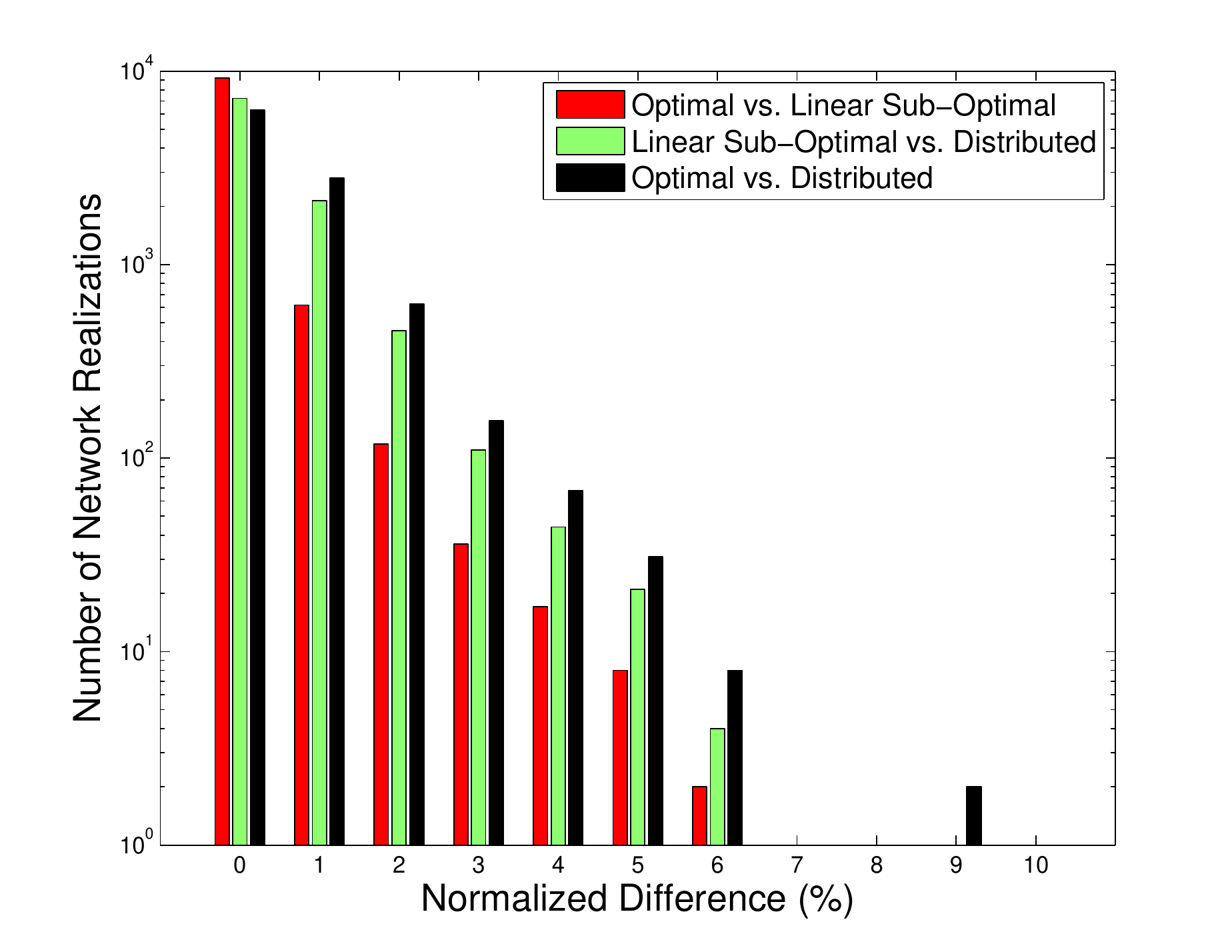}
     \caption{The histogram of the normalized differences between the energy consumptions of the proposed optimal, linear sub-optimal and distributed algorithms for $\lambda=0.03$.}
\end{figure}

\begin{figure}[h!]

  \centering
    \includegraphics[width=6in]{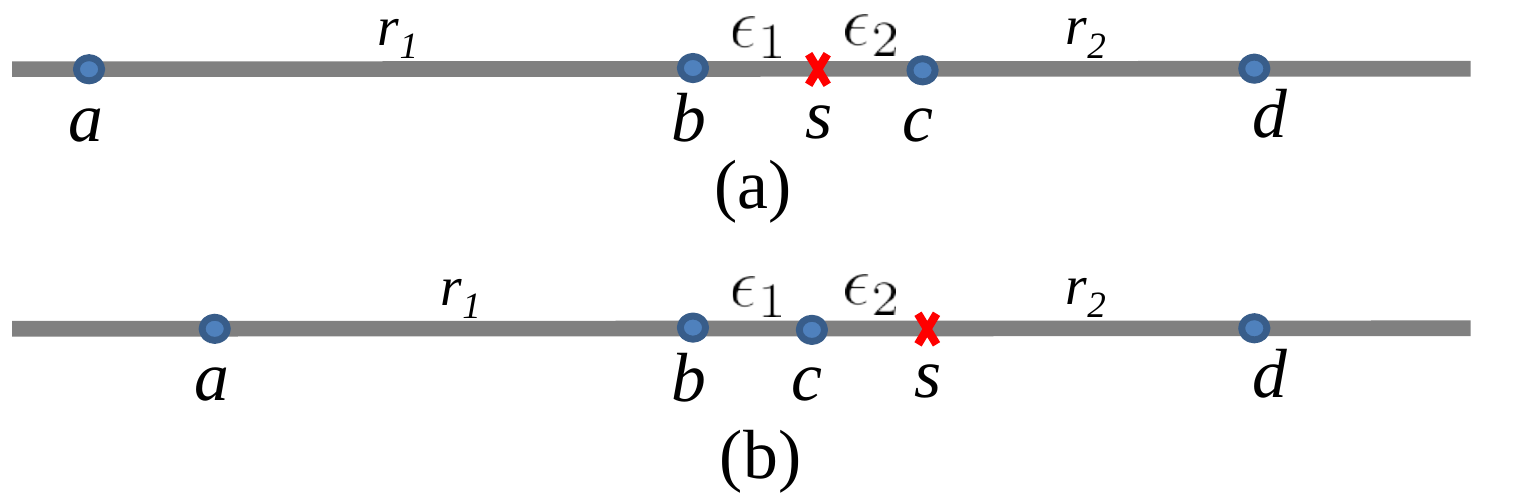}
     \caption{Network topologies where the difference between the consumed energy of different algorithms (linear sub-optimal vs. distributed in (a), and optimal vs. linear suboptimal in (b)) can be large.}
\end{figure}

\begin{figure}[h!]

  \centering
    \includegraphics[width=6in]{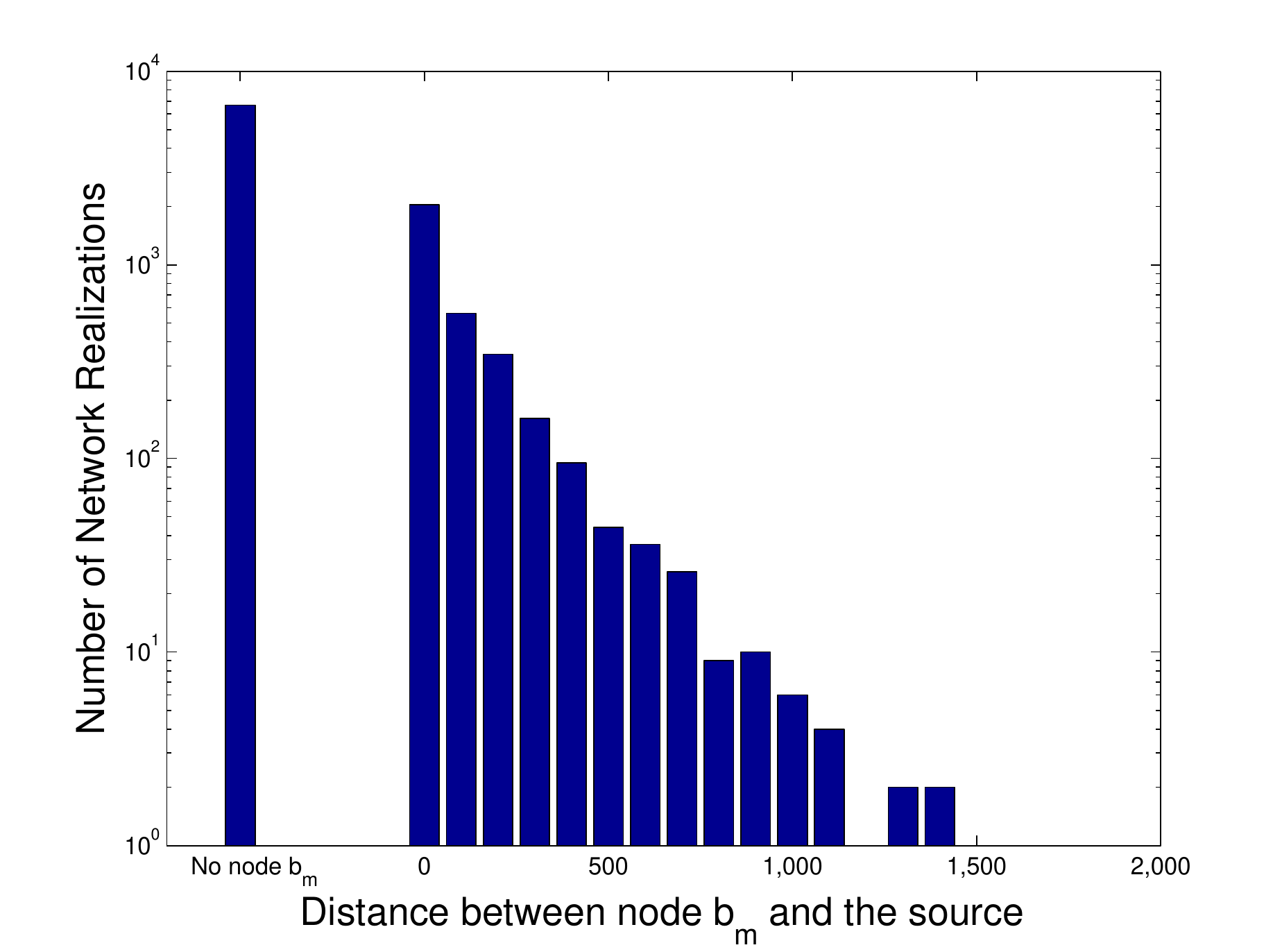}
     \caption{The histogram of the distance between node $b_m$ and the source node for $\lambda=0.03$.}
\end{figure}

\algsetup{
linenosize=\small,
linenodelimiter=.,
indent=1em
}

\begin{algorithm}

\caption{Sub-Optimal Linear-Time Algorithm}

\label{algRectangular}

\begin{algorithmic}[1]
\STATE Determine the sets $\mathbb{L}$ and $\mathbb{R}$

\STATE For each node $i$, calculate its opposite side coverage value $Cov(i)=M(i)-d(s,i)$

\STATE $m_L=\arg \max_{k}\left \{Cov(k)\right \}$ for $k \in \mathbb{L}$
\hskip 0.9em , \hskip 0.9em
$m_R=\arg \max_{k'}\left \{Cov(k')\right \}$ for $k' \in \mathbb{R}$

\STATE Denote by $l_{L}$ the left-most node in the opposite side coverage of $m_R$, and denote by $l_{R}$ the right-most node in the opposite side coverage of $m_L$

\STATE Calculate the costs:
\begin{align}
cost_R &=
\left\{
  \begin{array}{ll}
    cost^* \hskip 7em &\text{ if }l_R=s \nonumber\\
    \displaystyle\sum_{k=1}^{s_L}{M ^\alpha(k)} + \sum_{k=l_R}^{N}{M ^\alpha(k)} \hskip ,55em &\text{ if }l_R>s \nonumber
  \end{array}
\right.
\\\nonumber
\end{align}
\vspace{-1.5 cm}
\begin{align}
cost_L &=
\left\{
  \begin{array}{ll}
    cost^* \hskip 8em &\text{ if }l_L=s \nonumber\\
    \displaystyle\sum_{k=1}^{l_L}{M ^\alpha(k)} + \sum_{k=s_R}^{N}{M ^\alpha(k)} \hskip 1em &\text{ if }l_L<s \nonumber
  \end{array}
\right.
\end{align}
where
\vspace{-0.4 cm}
\begin{align}
cost^*=\max\{M(s_L),M(s_R)\}^\alpha + \underset{k\neq s}{\sum_{k=1}^{N}}{M ^\alpha(k)}\nonumber
\end{align}

If ($cost_R\leq cost_L$)
\vspace{-0.1 cm}
\begin{align}
R_{sub}(i)=
\left\{
  \begin{array}{ll}
    M(i) &i\in\{1,\ldots,s_L,l_R,\ldots,N\}\nonumber\\
    0  &i\in\{s_R+1,\ldots,l_R-1\}
  \end{array}
\right.
\end{align}
Else
\begin{align}
R_{sub}(i)=
\left\{
  \begin{array}{ll}
    M(i) &i\in\{1,\ldots,l_L,s_R,\ldots,N\}\nonumber\\
    0  &i\in\{l_L+1,\ldots,s_L-1\}
  \end{array}
\right.
\end{align}

Endif

\end{algorithmic}
\end{algorithm}

\algsetup{
linenosize=\small,
linenodelimiter=.,
indent=1em
}

\begin{algorithm}

\caption{Optimal Algorithm}

\label{algRectangular}

\begin{algorithmic}[1]
\STATE Construct arrays $C_S$, $C_E$ and the matrix $LR$

\STATE Perform Algorithm 1

\STATE $COST = Cost(R_{sub})$

\STATE
For each node $b\in \{1,\ldots,N\}$ do:

\hskip 1.5 em   4.1 Read nodes $rO_b$ and $rS_b$ from matrix $LR$

\hskip 1.5 em   4.2 $cost(b,rO_b,rS_b)=C_S[b]+[\max(d(b,rO_b),d(b,rS_b))]^\alpha+C_E[rS_b]+C_E[rO_b]$

\hskip 3.1 em    If $cost(b,rO_b,rS_b) < COST$

\hskip 4.5 em      $COST = cost(b,rO_b,rS_b)$, $b_m=b$, $rO=rO_b$ and $rS=rS_b$

\hskip 3.1 em    Endif

\hskip 1.5 em   4.3 Denote the next adjacent neighbors of nodes $rO_b$ and $rS_b$ by $nan(rO_b)$ and $nan(rS_b)$

\hskip 3.1 em    If either $nan(rO_b)$ or $nan(rS_b)$ exists

\hskip 4.5 em      Select the one which is closer to node $b$ (e.g. $nan(rS_b)$)

\hskip 4.5 em      Replace the element of $LR$ containing $rS_b$ with $nan(rS_b)$

\hskip 4.5 em      If{ $d(b,nan(rO_b)) = d (b,nan(rS_b))$}

\hskip 5.9 em    Replace the element containing $rO_b$ in $LR$ with $nan(rO_b)$

\hskip 4.5 em      Endif

\hskip 4.5 em    Goto 4.1

\hskip 3.1 em    Elseif  neither node exists and there are still some nodes left unprocessed

\hskip 4.5 em      Goto 4 with the next node $b$

\hskip 3.1 em    Else 

\hskip 4.5 em      Goto 5

\hskip 3.1 em    Endif

\STATE If no node $b_m$ has been found in 4

\hskip 1.3 em $R_{opt}=R_{sub}$

Else

\hskip 1.3 em $R_{opt}(b_m)=\max(d(b_m,rO),d(b_m,rS))$

\hskip 1.3 em $R_{opt}(i)=M(i)$ for all $i$ from $s$ up to $b_m$, and from $rO$ and $rS$ to both ends of the

\hskip 1.3 em  network correspondingly

\hskip 1.3 em $R_{opt}(k)=0$ for all the remaining nodes $k$

Endif

$cost(R_{opt})=COST$

\end{algorithmic}
\end{algorithm}

\end{document}